\documentclass[%
prab,
 amsmath,amssymb,
 reprint,
showkeys,
floatfix,
nofootinbib
]{revtex4-2}

\usepackage{graphicx}
\usepackage{dcolumn}
\usepackage{bm}
\usepackage{xcolor,soul}
\sethlcolor{green}
\usepackage[normalem]{ulem}
\usepackage[super]{nth}
\usepackage{easyReview}
\setreviewson

\usepackage[mathlines]{lineno}

\begin{document}

\preprint{AAPM/123-QED}

\title{Fast Particle-in-Cell simulations-based method for the optimisation of a laser-plasma electron injector}

\author{P. Drobniak }
\email{pierre.drobniak@ijclab.in2p3.fr}
\author{E. Baynard}
\author{C. Bruni}
\author{K. Cassou}
\author{C. Guyot}
\author{G. Kane}
\author{S. Kazamias}
\author{V. Kubytsky}
\author{N. Lericheux}
\author{B. Lucas}
\author{M. Pittman}
\affiliation{ 
Laboratoire de Physique des 2 Infinis Irène Joliot-Curie - IJCLab - UMR 9012 CNRS Universit\'e Paris Saclay - 91405 Orsay cedex - France 
}%

\author{F. Massimo}
\affiliation{%
Laboratoire de Physique des Gaz et des Plasmas - LPGP - UMR 8578, CNRS, Université Paris-Saclay, 91405 Orsay, France
}%

\author{A. Beck}
\author{A. Specka}
\affiliation{%
Laboratoire Leprince-Ringuet - LLR – UMR 7638 CNRS Ecole polytechnique, 91128 Palaiseau cedex – France
}%

\author{P. Nghiem}

\author{D. Minenna}
\affiliation{%
CEA-Irfu, Centre de Saclay, Université Paris-Saclay, 91191 Gif-sur-Yvette, France
}%

\date{\today}
\begin{abstract}

A method for the optimisation and advanced studies of a laser-plasma electron injector is presented, based on a truncated ionisation injection scheme for high quality beam production. The SMILEI code is used with laser envelope approximation and a low number of particles per cell to reach computation time performances enabling the production of a large number of accelerator configurations. The developed and tested workflow is a possible approach for the production of large dataset for laser-plasma accelerator optimisation. A selection of functions of merit used to grade generated electron beams is discussed. Among the significant number of configurations, two specific working points are presented in details. All data generated are left open to the scientific community for further study and optimisation.

\end{abstract}

\maketitle

\section{Introduction} \label{section:introduction} 

While conventional particle accelerators are getting larger and larger for high energy physics (27.6~km for the LHC \cite{bruning2012large} and 97.8~km for the FCC project \cite{benedikt2018fcc}), laser-driven wakefield acceleration (LWFA) is proving to be a promising technique for electron acceleration, yielding accelerating gradients three orders of magnitude greater than RF cavities, so in the range of 100~GV/m \cite{esarey1995trapping} thus requiring smaller facilities. Moreover, the mechanisms involved in the electrons injection tend to produce very short bunches, in the range of a few fs duration \cite{faure2009acceleration}. These characteristics make laser-plasma acceleration an interesting candidate for a new range of applications, such as electron sources for VHEE Flash therapy \cite{labate2020toward} and X-ray Free Electron Lasers (XFEL) \cite{oumbarek2021high}.

Recent characteristics of electron bunches experimentally generated by LWFA lie in the range of a few hundreds of MeV \cite{jalas2021bayesian}, \cite{kirchen2021optimal} up to a few GeV \cite{gonsalves2019petawatt}, with pC \cite{golovin2015tunable} up to nC \cite{couperus2017demonstration} charge, at a repetition rate around 1~Hz \cite{albert20212020}. They display a few percent energy spread \cite{albert20212020}, a normalised trace emittance around 1~mm.mrad \cite{kirchen2021optimal} and a divergence within the mrad range \cite{kirchen2021optimal}, \cite{lee2018optimization}. Note that these parameters are not all achieved simultaneously.

Physical mechanisms driving the injection and acceleration processes in LWFA for laser-plasma injectors (LPI) are highly non-linear and involve multiple coupled input parameters from the laser characteristics (focal spot position and size, focal distance, pulse duration and energy, polarisation, wavelength, spectrum) to plasma target parameters (gas choice, gas mixture composition, density distribution). Theoretical results and experimental demonstration allow for the rough choice of plasma density profiles and laser parameters \cite{couperus2018optimal} in order to achieve a desired electron beam. Nevertheless, these scaling laws are usually not sufficient to precisely simulate the tuning and optimisation of a laser-plasma accelerator (LPA). Moreover, due to high non-linearity of the coupled processes and the experimental difficulty to accurately measure and store shot-to-shot fluctuations, the stability around optimal injection and acceleration configurations is a critical point.

Therefore, utilising Particle-in-Cell (PIC) code, along with high performance computing resources and optimisation algorithms, has proven to be a valuable tool in LPA design and active control studies. Bayesian optimisation was already used and combined with experiments to deliver electron bunches at 1~Hz, with 250~MeV energy, subpercent energy spread and spectral density of $4.7$~pC/MeV \cite{jalas2021bayesian}. To our knowledge, massive generation of configurations (several thousands) in a short simulation time (a few hours), allowing the study of input-output correlations and with results open to the accelerator community has not been carried out yet.

The objective of this paper is to present a method for generating a large amount of PIC simulation results in a short time ($120$~core.hour $\approx{} 30$~minutes on 240 CPU-cores), using high performance computing (HPC) resources with moderate total computational costs. The generated results are useful in multiple aspects. They will first allow for the discovery of specific working points, displaying interesting characteristics for the injector. These specific working points can later on be better assessed by finer PIC simulations and also investigated regarding their stability. Finally, all generated beams can serve as input for building surrogate models using machine learning techniques to predict beam parameters. 

The choice of laser driver input parameters and plasma target configurations is defined in section~\ref{section:lpiparameters}. Then the numerical setup for fast simulations recently allowed by the PIC code SMILEI \cite{derouillat2018smilei} is presented and the massive random scan settings are introduced. An overview of the generated dataset is given where correlations between plasma target input parameters and electron beam output parameters are highlighted. Several possible functions of merit to quickly grade and compare the generated beams are discussed. Finally two different types of LPI configurations generating specific electron beams are extracted and further discussed. 

The results presented in this article are part of the PALLAS \cite{pallas} project at IJCLab, which uses the $1.6$~J moderate energy and $10$~Hz repetition rate laser provided by the LaseriX platform \cite{laserix}. PALLAS aims at optimising a LPI for the EUPRAXIA project \cite{eupraxia}, producing electron beams within the $150-250$~MeV energy range, less than $5$~\% energy spread, more than $30$~pC charge and a normalised phase emittance of less than $2$~$\mu$m \footnote{the beam divergence optimisation is out of the scope of the present study}. 

In the following, will be referred as '\textit{filter}' the condition '$Q >$ 30~pC \& $E_{med} >$ 150~MeV \& $\delta{}E_{mad} <$ 5~\% \& $\epsilon_{y,n} <$ 2~$\mu$m', where $Q$ is the charge, $E_{med}$ the median energy, $\delta{}E_{mad} = \sigma{}_{mad}/E_{med}$ (with $\sigma{}_{mad}$ the median absolute deviation) and $\epsilon_{y,n}$ the normalised phase emittance in y-direction (laser polarisation direction) defined as $\epsilon_{y,n} = \frac{1}{m_{0}c}\sqrt{\langle y^2 \rangle \langle p_{y}^2 \rangle - \langle yp_{y} \rangle^2}$ \cite{li2019preserving} (with $m_{0}$ the electron mass, $c$ the speed of light in vacuum and $p_{y}$ the momentum in y-direction).


\section{LPI parameters} \label{section:lpiparameters}

Experimental laser driver characteristics provided by LaseriX are: linearly polarised 5-th order Flattened Gaussian Beam (FGB) \cite{gori1994flattened} with $810$~nm central wavelength, spectral width of $\Delta \lambda = 30$~nm, a maximum energy on target of $1.6 \pm 0.1$~J, $35\pm 5$~fs duration ($10.5\pm 1.5$~$\mu{}$m length) at $10$~Hz. This corresponds to a peak power of $40$~TW. The laser-driver beam is focused with $1.5$~m focal length off axis parabola to a waist of $w_{0} = 19\,\mu$m. This leads to a laser intensity $a_{0}$ reaching its maximum in vacuum $a_{0,vac,max} = 1.40$ ($4 \times 10^{18}$~W.cm$^{-2}$). A laser upgrade could lower the pulse duration to $30$~fs and increase the energy to $2.4$~J, yielding an intensity of $a_{0,vac,max} = 1.85$. The Rayleigh length is $x_{R} = 1.42$~mm.

For the present electron density (see end of section \ref{section:lpiparameters}), the laser intensity is too low for self-injection (similar range studied in \cite{kuschel2018controlling}). So a nitrogen ionisation injection scheme \cite{pak2010injection} is chosen, using helium as main gas, in a $He+N_{2}$ mixture.
The plasma self-focusing in the target allows for an increase in $a_0$, high enough to ionise the inner shell electrons of nitrogen and potentially inject them in the wake. Indeed, the barrier suppression ionisation (BSI) potentials of $N_{2}$ two last electrons at $800$~nm \cite{couperus2018optimal} are $a_{0,BSI,N^{5+} \rightarrow N^{6+}} = 2.21$ and $a_{0,BSI,N^{6+} \rightarrow N^{7+}} = 2.77$. 

\begin{figure}[ht]
  \includegraphics[width=\linewidth]{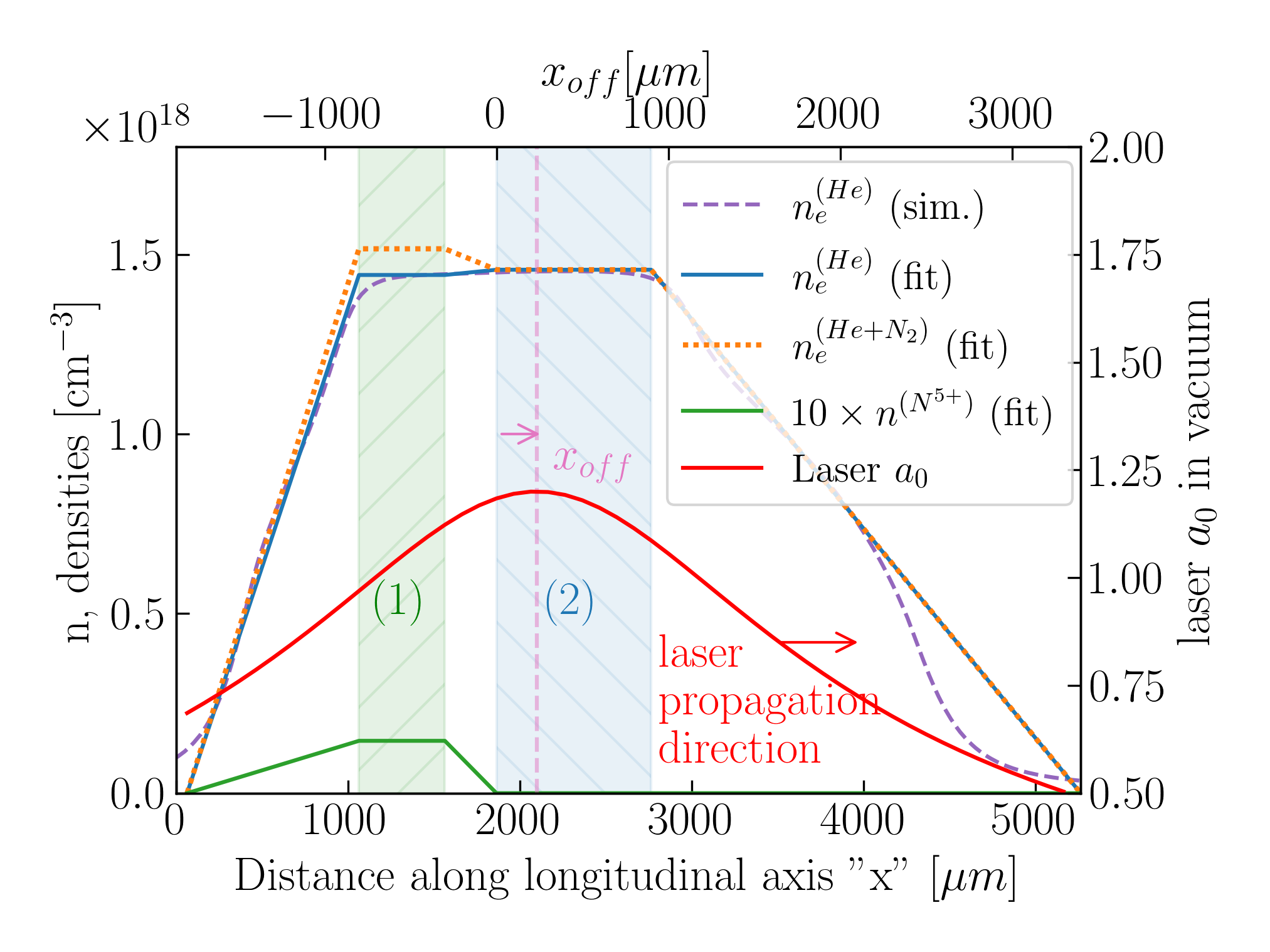}
  \caption{Example of electron density and dopant ion density for a target with $p = 30$~mbar and $c_{N_{2}} = 1\%$. $1^{st}$ and $2^{nd}$ plateaus respectively correspond to chamber 1 and chamber 2 and are delimited with dashed areas, respectively green and blue. $n_{e}^{(He)} (sim.)$ is the helium electron density, as simulated by OpenFOAM and assuming full ionisation and $n_{e}^{(He)} (fit)$ is the corresponding polygonal fit. $n_{e}^{(He+N_{2})} (fit)$ is the electron density fit on the mixture 'He+N2', assuming full ionisation of He and ionisation of $N_{2}$ up to the \nth{5} level. The atomic density of $N^{5+}$ ions is $n^{(N^{5+})}$. The envelope of a laser pulse traveling from left to right with maximum intensity in vacuum $a_{0,vac,max} = 1.2$ and focal position in vacuum $x_{off} = 235$~$\mu$m is added ($x_{off}$ reference starts at chamber 2 beginning).
            }
  \label{fig:laserAndDensity}
\end{figure}

\noindent As suggested in \cite{pak2010injection}, \cite{vargas2014improvements}, \cite{golovin2015tunable} and further investigated in \cite{lee2016dynamics} and \cite{audet2018gas}, the plasma target is split in two stages (Fig.~\ref{fig:laserAndDensity}). A first stage (chamber 1) with helium mixed with nitrogen (at molar concentration $c_{N_{2}}$ within a few percent range) dedicated to laser self-focusing and injection followed by a second stage (chamber 2) with helium only allowing for truncation of injection (no dopant anymore) and acceleration of the injected electron bunch. Since the two chambers share a common aperture (laser travels from chamber 1 to chamber 2), each chamber is set to the same pressure $p$ in order to prevent various species convection from one chamber to the other. The pressure $p$ then both describes the $He + N_{2}$ mixture pressure in chamber 1 as well as helium pressure in chamber 2. For a given pressure, the ratio between the electron densities of chamber 1 and 2 thus only comes from additional dopant concentration. The profile used here is generated by OpenFOAM \cite{openfoam} and a polygonal fit allows for a simplified direct variation of dopant concentration and overal pressure in the PIC simulations.

A first constraint on the choice of pressure is self-focusing. Based on the available laser power $P$, the corresponding electron density required for self focusing \cite{couperus2018optimal} is given by $P > P_c$ with $P_{c}[GW] \simeq 17 (\lambda_{p}/\lambda)^2$ where $P_{c}$ is the critical laser power required to trigger the effect, $\lambda{}_{p}$ and $\lambda{}$ respectively the plasma and laser wavelengths. Here the electron density necessary for self-focusing is $n_{e,sf} \approx 2.9 \times 10^{18}$~$cm^{-3}$. 
The pressure thus has to be greater than the self-focusing pressure $p > p_{He,sf} \approx 60$~mbar (assuming full ionisation of helium).

\noindent A second constraint on pressure is the optimal energy conversion between the laser driver and the plasma wave (resonant density). Based on the work done by Faure \cite{faure2009acceleration} and assuming a linear regime, one has to insure $k_{p} L_{0} = \sqrt{2}$, with $k_{p}$ the plasma wavenumber and $L_{0}$ the FWHM laser pulse length. In the present case, this yields an optimal electron density of $n_{e,opt} \approx 5.1 \times 10^{17}$~$cm^{-3}$ with corresponding fully ionised helium optimal pressure $p_{He,opt} \approx 10$~mbar.

\noindent Thus a helium pressure starting at $p_{He,opt}$ slightly lower than $p_{He,sf}$ is chosen, so a range $[10,100]$~mbar. Due to the experimental gas injection design (not presented in this article), pressures higher than $100$~mbar are not investigated.

With the choice of laser waist $w_{0} = 19\,\mu$m within the pressure range mentioned above, the focal spot is roughly matched for a potential bubble regime. Indeed, one can use the formula given by \cite{couperus2018optimal}: $w_{0} \simeq R_{b} = 2c\sqrt{a_{0}}/\omega_{p}$ where $R_{b}$ is the bubble radius and $\omega_{p}$ the plasma frequency. Using the extreme possible values for pressure and laser $a_{0}$, the corresponding matched spot size range is $[4.5,20.5]$~$\mu$m, so in agreement with the laser waist $w_{0} = 19\mu$m in vacuum (actually reaching even lower values due to self-focusing).

The depletion length $L_{pd}$ and dephasing length $L_{d,bubble}$ (if matched spot size) are defined as \cite{couperus2018optimal} $L_{pd} \simeq ({\omega_{0}/\omega_{p}})^{2} c \tau$ and $L_{d,bubble} \simeq  (2/3) \cdot (\omega_{0}/\omega_{p})^{2} w_{0}$ with $\omega{}_{0}$ the laser frequency, $\omega{}_{p}$ the plasma frequency, $\tau{}$ the laser duration. This yields $n_{e} = 5 \times 10^{18}\,$cm$^{-3}$ (dimensioning case), $L_{pd} \simeq 3.65$~mm and $L_{d,bubble} \simeq 4.41$~mm, well above the Rayleigh length $x_{R}$. Therefore $x_{R}$ is used for dimensioning the accelerating stage of our plasma target (chamber 2), so approximately $1$~mm.

The current experimental design for PALLAS project allows to vary four input parameters:
\begin{itemize}
	\item pressure $p \in{} [10;100]$~mbar
	\item laser $a_{0,vac,max} \in [1.1;1.85]$ (upper boundary to account for later laser upgrade)
	\item dopant concentration $c_{N_{2}} \in{} [0.2;12]\%$, defined as partial pressure ratio of dopant to mixture pressure ($0.2$\% is the minimum that can be experimentally achieve here, the upper value is based on previous work \cite{jalas2021bayesian})
	\item focal position offset $x_{off} \in{} [-400;1800]\mu$m, with origin $x_{off} = 0$ defined as focus position in vacuum and provided on upper horizontal axis of Fig.~\ref{fig:laserAndDensity} (inspired by \cite{couperus2018optimal}, \cite{jalas2021bayesian} and \cite{kirchen2021optimal}).
\end{itemize}

\section{Particle-in-Cell Simulations setup}

Simulations of  electron injection and acceleration in the plasma have been performed with the open source Particle-in-Cell (PIC) code SMILEI \cite{derouillat2018smilei,beck2018}. 

The physical setup assumes a laser propagation in $x$-direction and transverse plane on $y$- and $z$-axis. In case of cylindrical coordinates, space variables are $(r,\theta,x)$, with $x$ the laser propagation direction.

To speed-up LWFA simulations, which typically have a considerable computational cost, an envelope model \cite{Massimo2019,Benedetti2010} was used in cylindrical geometry, with only one azimuthal mode. Indeed, the coupling of cylindrical symmetry and envelope approximation can greatly reduce PIC simulations computational costs for LWFA, as shown in \cite{Benedetti2010,Tomassini2017,Massimo2019cylindrical}. The theoretical formula for FGB propagation was implemented (based on theory \cite{Santarsiero1997} and FBPIC implementation \cite{FBPIC_FGB}) and the laser was modeled as a \nth{5}-order FGB, with a waist of $w_{0} = 19\,\mu$m, $35$~fs FWHM unchirped Gaussian temporal profile and  $a_{0,vac,max}$ in [1.10,1.85]. 
Each simulation ran on 5 compute nodes with a total of $240\,$CPU-core (10 MPI processes each using 24 OpenMP threads).

Helium macro-particles were initialised fully ionised, while nitrogen macro-particles were initialised ionised up to the 5 first levels, both with initial temperature equal to zero (cold plasma). This approximation is justified by the fact that all helium electrons and the five first electrons of nitrogen are already ionised $50$~fs ahead of the pulse center, so approximately $1.5 \times T_{laser, FWHM}$, with $T_{laser, FWHM}$ the FWHM laser pulse duration. The main part of the pulse will thus propagate in an already ionised plasma of $He^{2+}$ and $N^{5+}$.

A moving window is used to follow the laser pulse in its propagation and keep only the physics of interest inside the simulation domain. Its characteristics are defined in 2D, with size in the ($x,r$) space set to $64~\rm{\mu m} \times 143~\rm{\mu m}$ ($6.10L_{0} \times 7.53w_{0}$), and a resolution of $\Delta x=0.1~\rm{\mu m}$ and $\Delta r=0.16~\rm{\mu m}$ with an integration time-step of $\Delta t=0.8 \Delta x / c = 0.27$~fs. The laser pulse center is in the simulation window, located $1.25 \times T_{laser, FWHM}$ from the window front edge, since $He^{2+}$ and $N^{5+}$ are already ionised at $1.5 \times T_{laser, FWHM}$.

For the simulation diagnostics, the electrons from He, the electrons from the five first levels of $N_{2}$ and N$_{2}$ inner shell electrons "born" from tunnel ionisation were tracked separately. This choice allowed to check which electrons came from ionisation injection and which from other injection mechanisms, e.g. downramp injection \cite{Esarey2009}. 

The electron density profile is read as input by the solver, as described in Fig.~\ref{fig:laserAndDensity}, where a polygonal fit on the OpenFOAM-simulated electron profile was performed, with space dimensions kept constant and where only the value of electron densities in plateau 1 and 2 were varied (through $p$ and $c_{N_{2}}$). 

For each species, only $1\,$macro-particle per cell ($ppc$) is used. The validity of such an approximation was checked by running a low and a high charge case respectively injecting $30$~pC and $160$~pC and comparing them with $8~ppc$ cases. The relative maximum error was $1\%$ on $E_{med}$, $10\%$ on $\delta{}E_{mad}$, $10\%$ on $Q$ and $8\%$ on $\epsilon_{y,n}$, which is acceptable for typical experimental measurement precision on these parameters.

The gain in computation time given by the reduction of $8$ to $1\,$ppc is significant. For low charge case, the computation time went from $450$~core.hour ($8$~ppc case)  to $130$~core.hour to ($1$~ppc case), so a speedup of $\times 3.5$. For high charge case, simulations were $\times 4$ faster from $700$~core.hour to $170$~core.hour for $8$~ppc and $1$~ppc cases respectively. So each scan simulation is performed with $1~$ppc and the average simulation time is approximately 30 minutes on 240 CPU-cores (time depends on the injected charge and the resulting number of macro-particles to track). Each configuration directory weighs around $5$~GB.

\section{Scan settings} \label{section:scansettings}

Using the GENCI high performance computing facility Irene-Joliot Curie \cite{GENCI}, five massive random scans (RS) called RS1, RS2, RS3, RS4 and RS5 in the ($p$, $a_{0,vac,max}$, $x_{off}$, $c_{N_{2}}$)-space were performed, with 2401 configurations each, so a total of 12005 simulations. Each RS ran for approximately 4 hours (limitation due to the maximum number of jobs authorised in the queue) and generated around $10$~TB of data. 
The input parameter space explored is presented in Table~\ref{tab:parametersRS}, where $SND_{900}$ and $SND_{1200}$ are \textit{Skew Normal Distributions} \cite{SND}, respectively centered around 900~$\mu$m and 1200~$\mu$m and all other parameters ranges follow random distributions.

\begin{table}[ht]
    \centering
    \begin{tabular}{m{1.4cm}|m{1.2cm}|m{1.3cm}|m{1.4cm}|m{1.4cm}|m{2cm}}
      & RS1 & RS2 & RS3 & RS4 & RS5 \\
    $p [mbar]$         & [10;100]      & [10;90] & [10;60] & [30;100] & [10;100] \\
    $a_{0,vac,max}$              & [1.1;1.45]    & [1.1;1.45] & [1.4;1.85] & [1.1;1.45] & [1.1;1.45] \\
    $x_{off}$ [$\mu$m]  & $SND_{900}$       & $SND_{1200}$ & [800;1800] & [800;1800] & [-400;600] \\
    $c_{N_2} [\%]$        & [0.2;12]      & [0.5;2] & [0.5;12] & [0.5:2] & [0.5:2]
    \end{tabular}
    \caption{Input parameters investigated for RS1, RS2, RS3, RS4 and RS5. $SND_{900}$ and $SND_{1200}$ are \textit{Skew Normal Distributions} \cite{SND}, respectively centered around 900~$\mu$m and 1200~$\mu$m. Other parameters are picked randomly within the specified range.}
    \label{tab:parametersRS}
\end{table}

Random scans make easier the visualisation of the 4D-input space on 2D or 3D meshes since points do not overlap. They also allow for randomly distributed small variations of input parameters on which the output might be very sensitive. Non-deterministic randomised combinations of the hyper parameter input space have been generated (see git repository for more information \cite{git}).

\section{Results}

\subsection{Post-processing}  \label{subsection:postprocessing}

A python script based on HAPPI library \cite{happi} is used for post-processing to extract the electron beam and laser parameters \cite{git}. 
For the injection the inner shell electrons from $N^{5+}$ were tracked. The electron beam data are extracted at the last simulation time step (end of the plasma density out ramp) and a lower cut-off energy of $25$~MeV is applied on the electron bunch energy distribution. Low charge beams below $0.3$~pC are not considered. Electrons originating from Helium are not included in the resulting beam, since they are very rarely trapped by self injection ($a_{0} > 4$) and do not contribute to the overall charge.

All post-processing scripts are available online \cite{git}. 

\subsection{Injection conditions} \label{subsection:injectionConditions}

\begin{figure}[ht]
  \includegraphics[width=\linewidth]{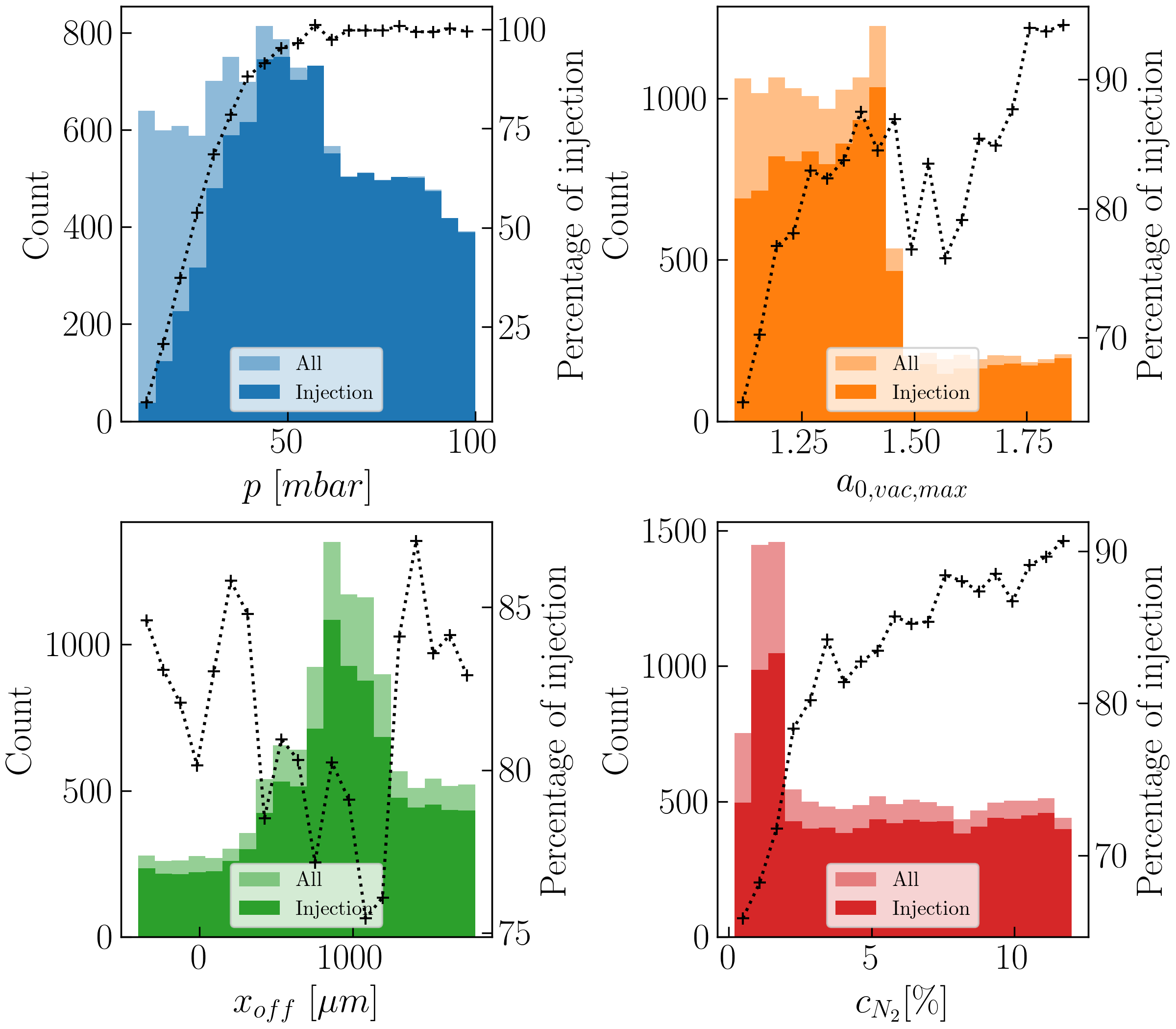}
  \caption{Histogram comparison between injection and non-injection for all RS (bins $= 20$). All configurations are displayed in light colours, the ones leading to injection are in dark colours. For each bin, the percentage of injection is represented by black crosses.}
  \label{fig:injectionHist}
\end{figure}

The scans RS1, RS2, RS3, RS4 and RS5 show effective injection (i.e. the integrated charge above $25$~MeV must be superior to $3$~pC) respectively in 80\%, 66\%, 83\%, 92\% and 82\% of cases, so a total of 10025 generated beams. RS2 tried very downstream focuses so injected less than other RS. RS4 did not try very low pressures, so ensured very often self-focusing. The maximum effective $a_{0,eff,max}$ reached within propagation was high enough for ionising $N_{2}$ inner shell electrons and for generating a large bubble, thus favouring ionisation injection. The injection triggering trend for each input parameters is summarised in Fig.~\ref{fig:injectionHist}, where all configurations tried in the ($p$, $a_{0,vac,max}$, $x_{off}$, $c_{N_{2}}$)-space are displayed in light colours while the ones leading to injections are in dark ones.

Injection triggering is favoured by high $p$ (stronger self-focus), high $a_{0,vac,max}$ (higher tunnel ionisation rate) and high $c_{N_{2}}$ ($N_{2}$ participates in the background electron density). An upstream $x_{off}$ means high intensity while entering chamber 1 so strong self-focusing and thus high tunnel ionisation rate in the dopped region. A downstream $x_{off}$ has lower intensity while entering chamber 1 and a maximum reached later in the propagation. If this maximum happens at the very end of chamber 1, injection in this case is low. The unexpected rise for $x_{off}$ above $1000~\mu$m is explained by the hidden $a_{0,vac,max}$ parameter, which is higher in the case of RS3.

\begin{figure*}[ht]
  \includegraphics[width=\linewidth]{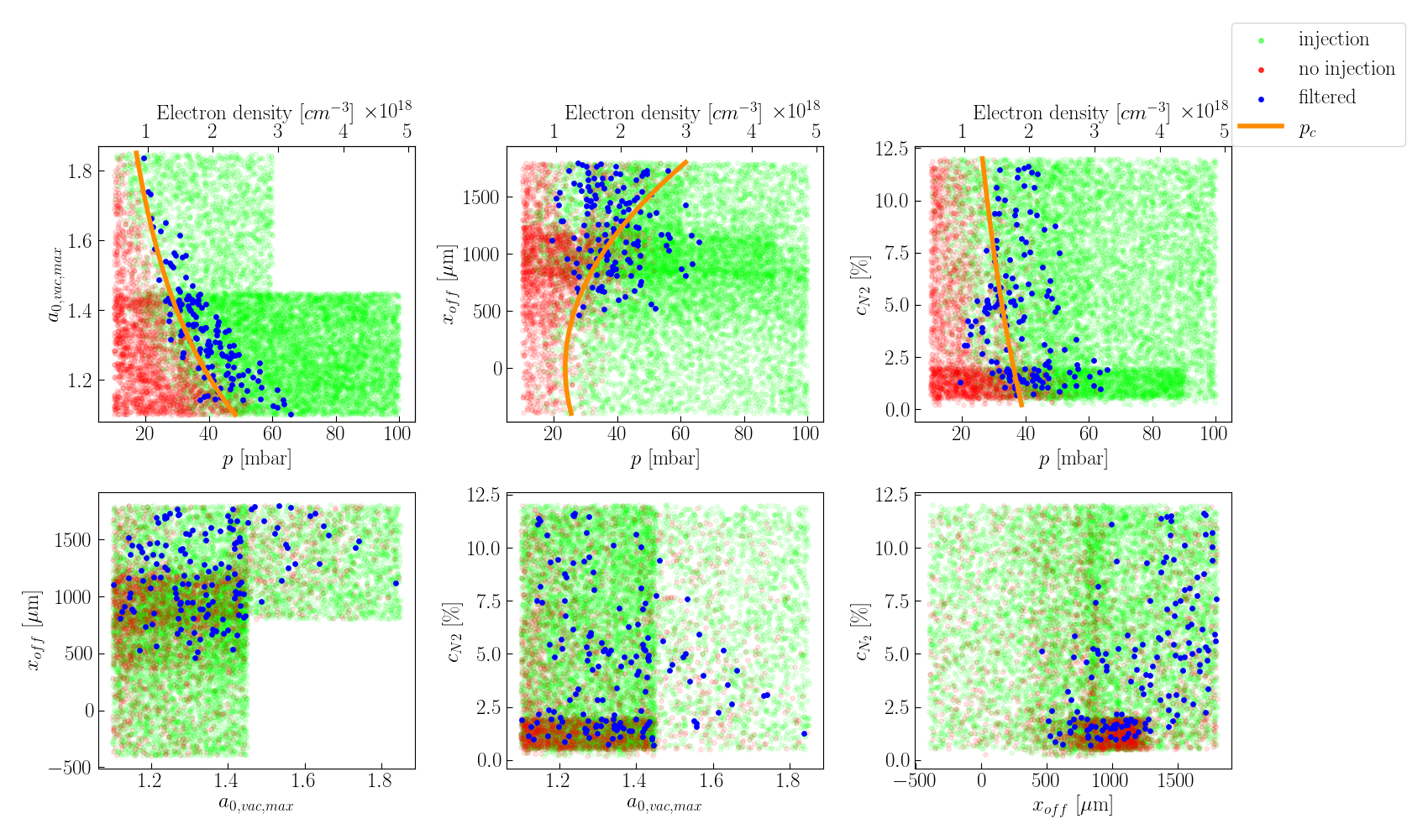}
  \caption{Comparison between injection (green dots) and non-injection (red dots) for all configurations in the input space. Critical pressure (or electron density) required for self-focusing is computed using the formula for self-focusing given in \cite{lu2007generating}, where hidden parameters $x_{off}$ and $c_{N_{2}}$ in the ($p,a_{0}$)-graph, $a_{0}$ and $c_{N_{2}}$ in the ($p,x_{off}$)-graph and $a_{0}$ and $x_{off}$ in the ($c_{N_{2}}$)-graph were averaged. Configurations satisfying the filter defined in section~\ref{section:introduction} are displayed with blue dots.}
  \label{fig:injectionScatter}
\end{figure*}

Cross correlations on input parameters triggering injection are presented in Fig.~\ref{fig:injectionScatter}, where configurations producing the target electron beam (satisfying the filter defined in section~\ref{section:introduction}) were identified in dark blue.

Using the formula for self-focusing given in \cite{lu2007generating}, the \textit{'mean'} critical pressure $p_{c}$ required for the laser self-focusing in the plasma target first chamber is added through the yellow curve. The term \textit{'mean'} refers to an average on the two hidden parameters of each 2D plot.

In the ($p,a_{0,vac,max}$)-graph of Fig.~\ref{fig:injectionScatter} one can see a correlation for injection between $p$ and $a_{0,vac,max}$ and a diffuse transition between the \textit{"no-injection"}-area to the \textit{"injection"}-area, well fitted by the $p_{c}$ curve. As a matter of fact, injection needs self-focusing to appear and since there are two hidden parameters, this transition is diffuse. The target beams lie at the transition to injection, because too much self-focusing will induce injection in a large volume (longitudinally and transversally) and thus produce too much charge leading to a beam with too high energy spread, emittance and possibly too low energy (beam-loading). One can also note that the filtered beams can be produced within all the tested $a_{0,vac,max}$ range, while $p$ has to be kept within $[20;70]\,$mbar.

In the ($p,x_{off}$) view, a correlation for triggering of injection also appears between $p$ and $x_{off}$,  fitted by $p_{c}$, since an upstream focus means that the laser enters chamber 1 with a higher $a_{0}$, thus facilitating self-focusing. For very high $x_{off}$ two separate injection regions appear and there is no sharp transition by varying $p$. This is explained in the ($ a_{0,vac,max},x_{off}$) graph, where one sees that very high $a_{0,vac,max}$ were also tried for downstream focuses (contribution from RS3). Here again, the target beams lie at this diffuse transition, since they require a small injection volume. There is a preferred region in $x_{off}$ to produce them above $500\,\mu$m, since upstream focus will inject too much charge: space charge and beam loading effects will affect the energy, energy spread and emittance. Higher $x_{off}$ above $1800\,\mu$m are also interesting to generate the target beams, but seem to require always higher $a_{0,vac,max}$ which are out of the present study.

In the ($p,c_{N_{2}}$) graph one sees a threshold on pressure $p$ (required for self-focusing to happen) correlated with the dopant concentration $c_{N_{2}}$. The transition between \textit{"no-injection"} and \textit{"injection"} is well described by $p_{c}$. 

It is hard to identify particular trends in the ($a_{0,vac,max},x_{off}$), ($a_{0,vac,max},c_{N_{2}}$) or ($x_{off},c_{N_{2}}$) views, both for the injection points or for the target beams since the pressure $p$ plays a very significant role but is a hidden parameter in these three graphs. Still, one sees that focusing too upstream will not produce target beams. The reason for this lies in a strong self-focusing producing very high-charge beams displaying poor characteristics.

As a conclusion on the injection tendencies, a strong dependance on $p$ appears to trigger injection with a diffuse transition from \textit{"no-injection"} to \textit{"injection"} in the ($p$,$a_{0,vac,max}$), ($p$,$x_{off}$) and ($p$,$c_{N_{2}}$) graphs, since injection requires self-focusing to appear and this phenomenon is dependent on the electronic density (mostly $p$ but also $c_{N_{2}}$ through outer shell electrons) and the laser intensity at chamber 1 entrance, so $a_{0,vac,max}$ and $x_{off}$. This transition is well fitted by the theoretical curve $p_{c}$ for self-focusing. A control on the volume of injection is a critical point for the target beams.

\subsection{Electron beams evaluation}

The output space of interest for the present study is defined by the following electron beam parameters:  $(Q,E_{med},\delta{}E_{mad},\epsilon_{y,n})$. Divergence was not included since this parameter can be controlled by optimising the plasma out-ramp. It was already studied in \cite{dornmair2016dedicated} \cite{li2019preserving} and experimentally demonstrated \cite{dickson2022mechanisms}. All results are available online and the reader can use \add{the online dashboard} for their own data exploration \cite{git}.

In addition to the filter condition ($F = 1$ if the filter is fulfilled, $0$ otherwise), different functions of merits inspired by the litterature are tried \footnote{'functions of merit' should not be mistaken with 'objective functions', since the former allow for a scalar view of the output space, while the latter are used in a decision process (in a Bayesian optimisation process for example)} (eq.~\ref{eq:functionsOfMerit}):

\begin{equation} \label{eq:functionsOfMerit}
   \begin{gathered}
    f_{1} = \frac{E_{med}^2 \cdot Q}{\sigma{}_{mad} \epsilon_{y,n}} \\ 
    f_{2} = \frac{E_{med} \cdot \sqrt{Q}}{\sigma{}_{mad}}\\ 
    f_{3} = \frac{E_{med} \cdot  Q}{\sigma{}_{mad}}\\ 
    f_{4} = \frac{E_{med} \cdot Q}{\sigma{}_{mad} \sqrt{\epsilon_{y,n} \epsilon_{z,n}}} 
    \end{gathered}
\end{equation}   

There is no universal function of merit since each application requires the optimisation of given sets of beam characteristics:  $f_{1}$ gives more importance to $E_{med}$, $f_{2}$ is the function used by Jalas et al. in their Bayesian optimisation \cite{jalas2021bayesian}, $f_{3}$ insists more on charge, $f_{4}$ includes the normalised transverse phase emittances $\epsilon_{y,n}$ (laser polarisation direction) and $\epsilon_{z,n}$ to optimise the brightness.

\subsection{Best beams generated}

\begin{figure*}[ht]
  \includegraphics[width=\linewidth]{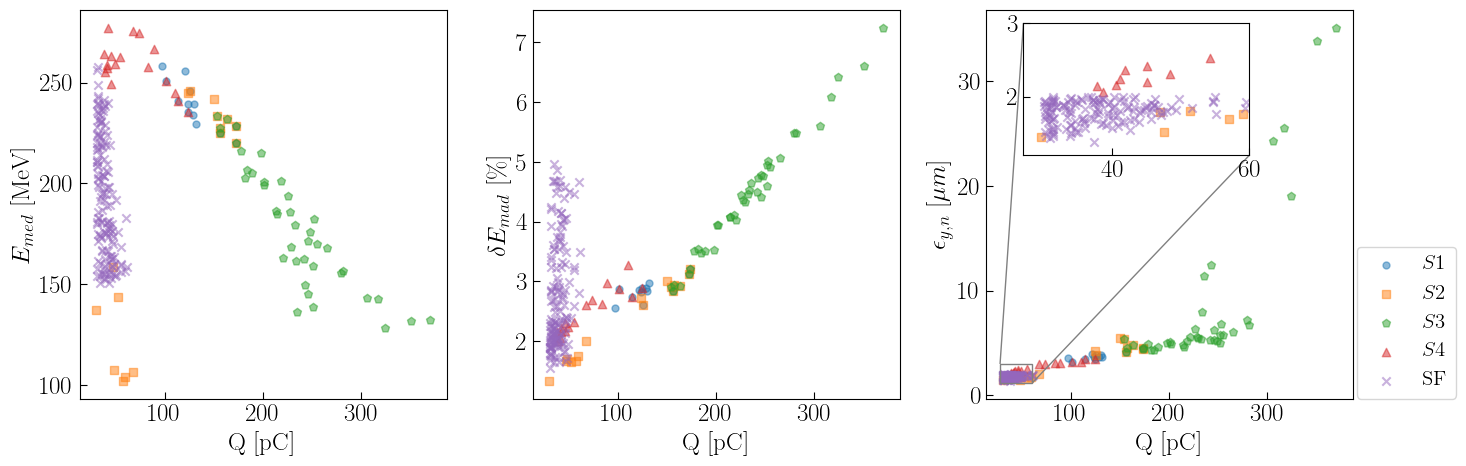}
  \caption{Three views of simulations results in the output space as function of the injected charge $Q$. Beams selected by the functions of merit with a cut-off at 90\% of $f_{1}$, $f_{2}$, $f_{3}$, $f_{4}$ maximum and beams in the filter defined in section \ref{section:introduction} are compiled in sets respectively denoted $S1$, $S2$, $S3$, $S4$ and $SF$.}
  \label{fig:bestConfigsOutputCharge}
\end{figure*}

Beams were selected to a cut-off of 90\% of each function maximum. Results are presented in Fig.~\ref{fig:bestConfigsOutputCharge}, where three views of the output space were chosen: $(Q,E_{med})$, $(Q,\delta{}E_{mad})$ and $(Q,\epsilon_{y,n})$. From now on, the terms $S1$, $S2$, $S3$, $S4$ and $SF$ are used to write about sets respectively selected by $f_{1}$, $f_{2}$, $f_{3}$, $f_{4}$ and $F$. 

In the $(Q,E_{med})$ view, a maximum appears for $E{med}$ for $Q$ within $[80,100]$~pC, followed by a linear decrease with $Q$. Such a behaviour is explained by beam-loading effects, where more charge flattens the longitudinal accelerating field. 

The $(Q,\delta{}E_{mad})$ graph shows a quasi linear increase of $\delta{}E_{mad}$ with $Q$, except for a specific region within $[80,150]$~pC, where a stagnation appears (optimal working point).

The $(\epsilon_{y,n},Q)$ graph also displays a linear increase of $\epsilon_{y,n}$ with $Q$, the slope being more pronounced for $S3$ which favours high charges. High charge leads to strong space-charge effects and a higher normalised transverse emittance. Furthermore, high charge beams are loaded even far from axis, inducing strong oscillations of the electrons in the transverse plane and thus high $\epsilon_{y,n}$. 

Looking at functions of merit, one can say that $f_{1}$ and $f_{4}$ are good compromises in terms of $Q$, $\delta{}E_{mad}$, $E_{med}$ and $\epsilon_{y,n}$, the latter remaining a bit too high for the beams to be in $SF$. $f_{2}$ favours two areas of the output space, so its use in a decision process might not be optimal for the present parameter range. $f_{3}$ is useful for highlighting very high charge beams, regardless of their $\epsilon_{y,n}$.

The results generated by all RS produced 145 configurations in $SF$. $f_{1}$, $f_{2}$ and $f_{4}$ managed to approach it, with only one configuration in $S2 \cup SF$. In this selection, the maximum $E_{med}$ was $257$~MeV and lowest $\delta E_{mad}$ reached $1.54\,\%$ ($\sigma_{mad} =3.27\,$MeV). The charge is very limited, since $Q$ above $61$~pC generated $\epsilon_{y,n}$ above the filter limit.

\begin{figure*}[!ht]
  \includegraphics[width=\linewidth]{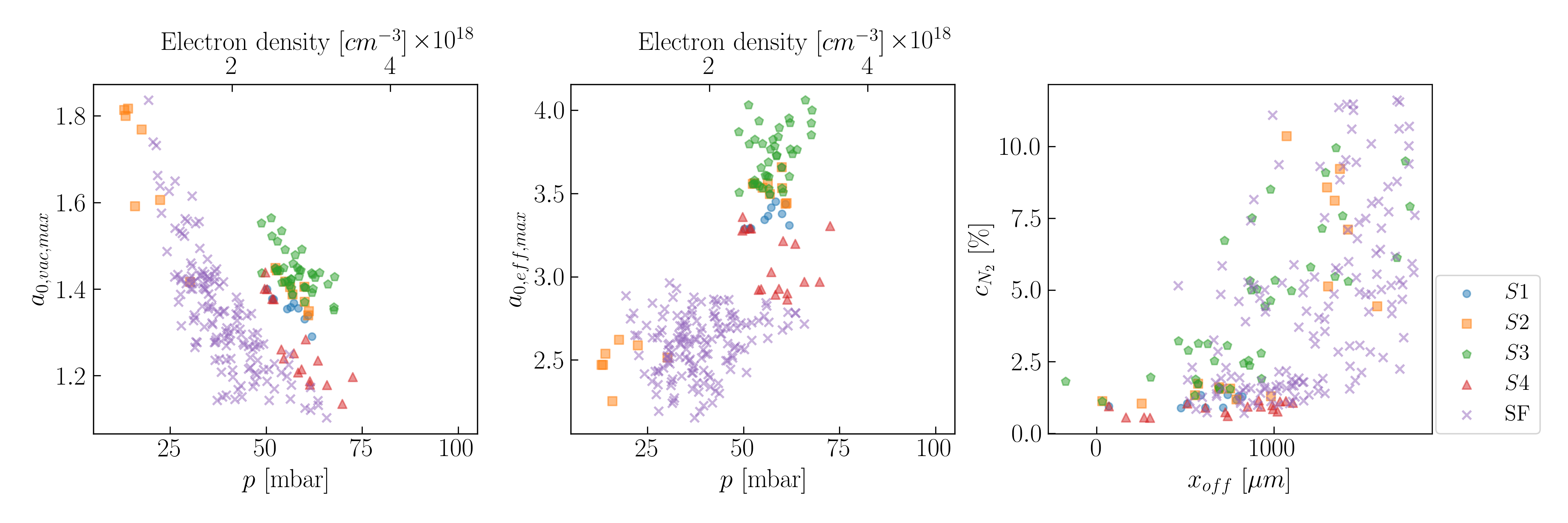}
  \caption{Three views of simulations results in the $(p,a_{0,vac,max},x_{off},c_{N_{2}})$ input space, where a plot with $a_{0,eff,max}$ was added, corresponding to the effective maximum laser intensity reached by the laser in the plasma. Beams selected by the functions of merit with a cut-off at 90\% of $f_{1}$, $f_{2}$, $f_{3}$, $f_{4}$ maximum and beams in the filter defined in section \ref{section:introduction} are compiled in sets respectively denoted $S1$, $S2$, $S3$, $S4$ and $SF$.}
  \label{fig:bestConfigsInput}
\end{figure*}

\subsection{Best injector configurations}\label{subsection:bestconfig}

The input space parameters of $S1$, $S2$, $S3$, $S4$ and $SF$ are presented in Fig.~\ref{fig:bestConfigsInput}. An axis with the variable $a_{0,eff,max}$ is added, which corresponds to the maximum of $a_{0}$ effectively reached by the laser during its propagation in the plasma.

As a first consideration on configuration distribution in Fig.~\ref{fig:bestConfigsInput}, one can see that $S1$, $S2$, $S3$, $S4$ and $SF$ all seem to gather in different areas of the $(p,a_{0,vac,max})$ and $(p,a_{0,eff,max})$-spaces, with a slight overlap between $S2$ and $S3$, as already seen in Fig.~\ref{fig:bestConfigsOutputCharge}. In the $(x_{off},c_{N_{2}})$-space, $S2$, $S3$ and $SF$ occupy large areas, which means that $x_{off}$ or $c_{N_{2}}$ were not the critical variables for each set. However, $S1$ and $S4$ occupy well defined areas, with low $c_{N_{2}}$, where $S1$ is precisely located around $x_{off} = 700\,\mu$m and $S4$ is more flexible on $x_{off}$.

High charge beams ($S3$) originated from the highest $p \times a_{0,vac,max}$ combination, resulting in the highest $a_{0,eff,max}$, above $3.5$. They were produced within a wide range of $x_{off}$ and $c_{N_{2}}$.

On the contrary, low charge beams (from $SF$ for instance) originated from relatively low $p$ and $a_{0,vac,max}$, inducing an $a_{0,eff,max}$ in the range of $N^{5+}$ and $N^{6+}$ BSI.

The optimal zone with highest $E_{med}$ in Fig.~\ref{fig:bestConfigsOutputCharge} $(Q,E_{med})$-space is described by $S4$ and corresponds to $p \in [50;70]$~mbar, $a_{0,vac,max} \in [1.1;1.4]$ and $x_{off} \in [0;1100]\,\mu$m leading to an $a_{0,eff,max} \in [2.9;3.3]$.

The stagnation area identified in Fig.~\ref{fig:bestConfigsOutputCharge}$(Q,\delta{}E_{mad})$-space described by $S1$ is reached for $p \approx 60$~mbar, $a_{0,vac,max} \approx 1.4$ and $x_{off} \in [500;700]\,\mu$m, leading to an $a_{0,eff,max} \approx 3.5$. These beams were obtained with very low $c_{N_{2}}$, below $2\%$.

Beams in $SF$ originate from a wide range of $p$ and $a_{0,vac,max}$ but in a certain $p \times a_{0,vac,max}$ area. The focus $x_{off}$ had to be downstream $500\,\mu$m. No particular constraint appears on $c_{N_{2}}$. By looking at the $(p,a_{0,eff,max})$-space, one sees that $a_{0,eff,max}$ has to remain within the range of $N^{5+}$ and $N^{6+}$ BSI intensities to trigger tunnel ionisation but not too strongly. This reduces the volume of injection either longitudinally (reduced injection length) or transversally (no injection far from axis).

\subsection{Selected configurations}

In this section, we analyse two LPI configurations originating from very different combinations of input parameters:
\begin{itemize}
    \item best of $f_{3}$ (config. 3702, comes from RS2)
    \item $\delta{}E_{mad}^{(min)}$ (lowest energy spread) in the filter (config. 7516, comes from RS4)
\end{itemize}

\begin{figure}[ht]
  \includegraphics[width=\linewidth]{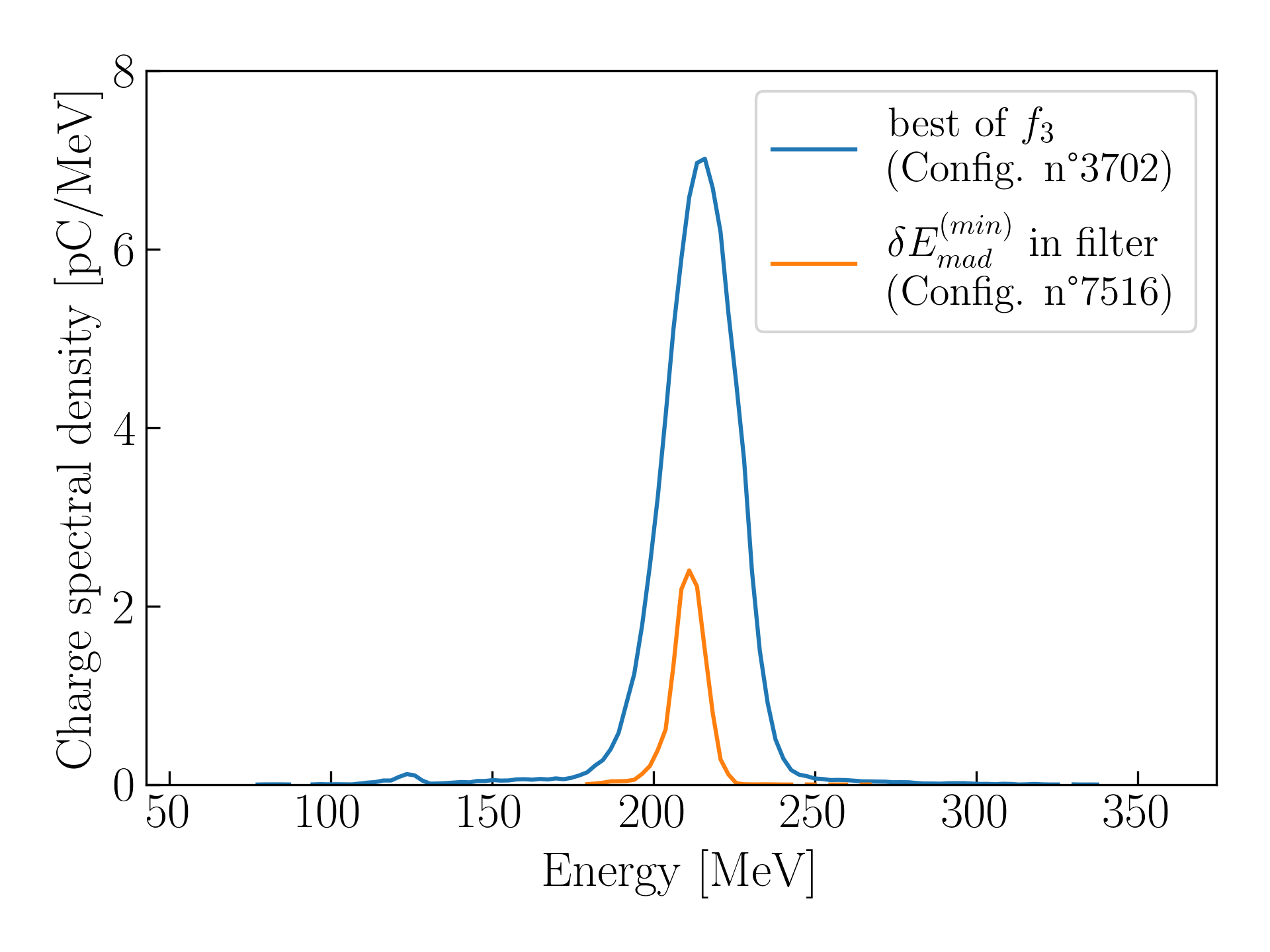}
  \caption{Spectra of the beams generated by the two simulations of configurations $3702$ and $7516$, respectively corresponding to the best of $f_{3}$ and $E_{mad}^{(min)}$ (lowest energy spread) in filter from section \ref{section:introduction}. Electrons below 25~MeV were not captured.}
  \label{fig:stability2Spectra}
\end{figure}

The spectra of those two beams are presented in Fig.~\ref{fig:stability2Spectra}.

The input and output parameters of these two configurations are presented in Table~\ref{tab:parametersStability}.
\begin{table}[ht]
    \centering
    \begin{tabular}{p{1.8cm}|p{2cm}|p{2.4cm}}
      & best of $f_{3}$ & $\delta{}E_{mad}^{(min)}$ in filter \\
      \hline \hline
      N$^\circ$ & 3702 & 7516 \\
      Origin & RS2 & RS4 \\
      \hline
      $p [mbar]$ & 58.6 & 47.8 \\
      $a_{0,vac,max}$ & 1.43 & 1.23 \\
      $x_{off} [\mu m]$ & 558 & 1680 \\
      $c_{N_{2}}$ [\%] & 1.88 & 6.17 \\
      \hline
      $a_{0,eff,max}$ & 3.73 & 2.58 \\
      \hline
      $Q [pC]$ & 198 & 30 \\
      $E_{med} [MeV]$ & 215 & 212 \\
      $\delta{}E_{mad} [\%]$  & 3.53 & 1.55 \\
      $\epsilon_{y,n} [\mu m]$ & 5.03 & 1.74
    \end{tabular}
    \caption{Input and beam parameters of LPI configurations 3702 and 7516.}
    \label{tab:parametersStability}
\end{table}

Configuration $3702$ corresponds to relatively high pressure ($p = 58.6$~mbar) and relatively strong intensity ($a_{0,vac,max} = 1.43$) at upstream focus ($x_{off} = 558\,\mu$m) leading to strong self-focusing ($a_{0,eff,max} = 3.73$), so inducing a very high injected charge ($Q = 198$~pC) even for a low dopant concentration ($c_{N_{2}} = 1.88\%$). This high charge induces a high emittance ($\epsilon_{y,n} = 5.03\,\mu$m) and does not fit $SF$. Configuration $7516$ was generated with moderate pressure ($p = 47.8$~mbar), relatively low intensity ($a_{0,vac,max}$ = 1.23), a downstream focus ($x_{off} = 1680\,\mu$m) leading to mild self-focusing ($a_{0,eff,max} = 2.58$) in a medium dopant concentration ($c_{N_{2}} = 6.17\%$), so a quite low injected charge ($Q = 30$~pC). This beam displays very small energy spread ($E_{mad} = 1.55\%$) and emittance ($\epsilon_{y,n} = 1.74\,\mu$m) while remaining in $SF$.

Beam $3702$ is interesting for high energy physics and FLASH therapy application, while beam $7516$ could be an interesting candidate for X-FEL generation due to its reduced energy spread. 

\begin{figure*}[ht]
  \includegraphics[width=12cm]{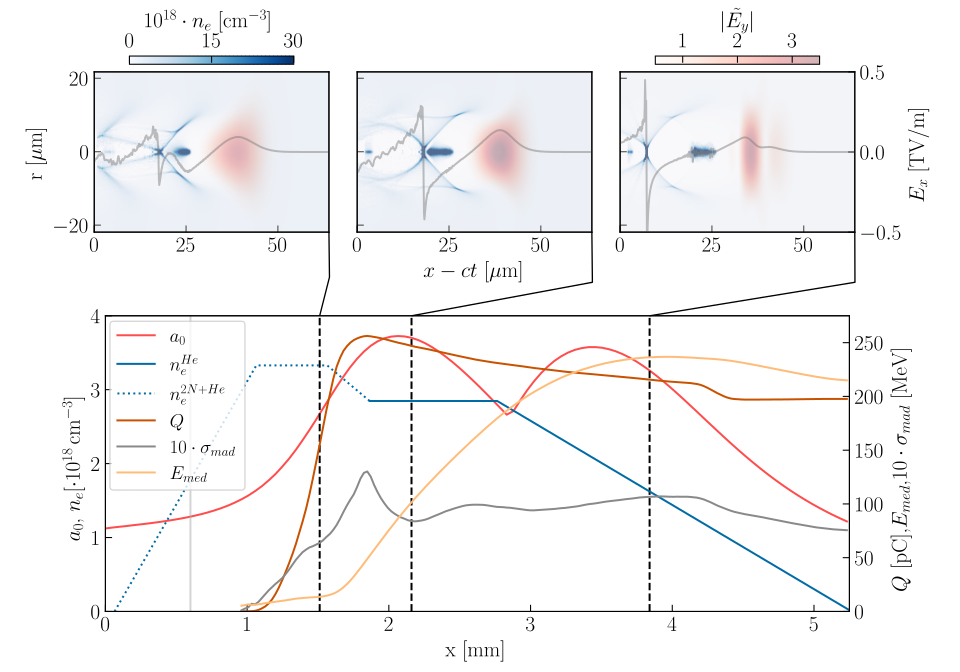}
  \caption{Evolution of laser ($a_{0}$) and beam parameters ($Q$, $E_{med}$, $\sigma{}_{mad}$) during propagation for configuration 3702. The electron density profile for chamber 1 (dashed line) and chamber 2 (solid line) is added. Laser travels from left to right. Three snapshots display the injection process at three different timesteps (increase of $a_{0}$, maximum of $a_{0}$ and beam at plasma outramp). Entrance of chamber 1 is located at $x = 1$~mm.}
  \label{fig:3702}
\end{figure*}

\begin{figure*}[ht]
  \includegraphics[width=12cm]{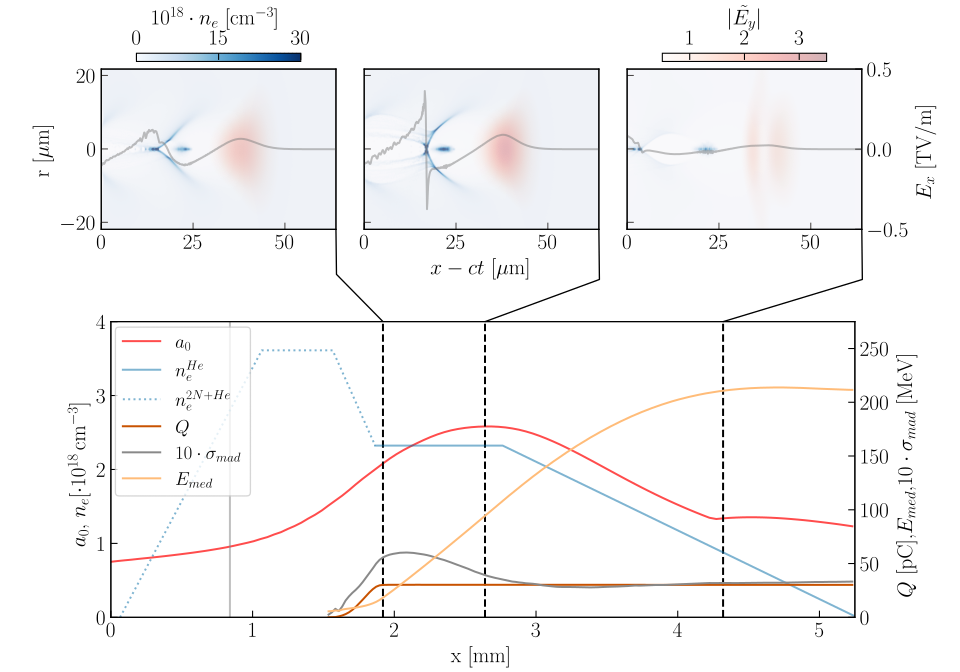}
  \caption{Evolution of laser ($a_{0}$) and beam parameters ($Q$, $E_{med}$, $\sigma{}_{mad}$) during propagation for configuration 7516. The electron density profile for chamber 1 (dashed line) and chamber 2 (solid line) is added. Laser travels from left to right. Three snapshots display the injection process at three different timesteps (increase of $a_{0}$, maximum of $a_{0}$ and beam at plasma outramp). Entrance of chamber 1 is located at $x = 1$~mm.}
  \label{fig:7516}
\end{figure*}

More details on the beam dynamics of configurations 3702 and 7516 are given in Fig~\ref{fig:3702} and Fig.~\ref{fig:7516}, where self-focusing occurs for configuration 3702. $a_{0}$ reaches its maximum upstream chamber 2 entrance, followed by laser guiding during propagation. An early injection starts in chamber 1 around $x = 1$~mm and lasts for almost $1$~mm, while the injection for configuration 7516 starts very late at $x = 1.8$~mm (typically the entrance of chamber 2) and stops $0.2$~mm later.

The charge $Q$ reaches its maximum at beginning of chamber 2 (zone where the dopant is no longer present) for both cases. For configuration 3702, one sees that $Q$ then decreases during propagation. This comes from the presence of a second bunch (whose charge is also accounted for in $Q$) behind the bubble which tends to slip out of the box during propagation.

Beam-loading is present for both configurations and even surprisingly for the lower charge case ($30$~pC). It is clearly visible on Fig.~\ref{fig:7516}, where the longitudinal accelerating field $E_{x}$ becomes almost constant along the bunch. For the higher charge case, the observed high value of $a_{0}$ (almost twice higher) mitigates the beam-loading effect (typical sharp bubble shape as observed in the laser-dominated regime \cite{gotzfried2020physics}).


For configuration 7516, the propagation in chamber 2 has a positive effect on the energy spread $\sigma{}_{mad}$, while its effect is less noticeable for configuration 3702. As a matter of fact, the longitudinal accelerating field $E_{x}$ is not constant along the bunch for configuration 3702 (beam-loading is not optimal).

As expected, chamber 2 (including the downramp) clearly plays its role for accelerating the bunch.

The careful analysis of these two configurations shows that the cell design is particularly relevant: chamber 1 allows for self-focusing triggering injection and chamber 2 is efficiently designed for energy increase and energy spread reduction. 

These two particular configurations show clear influence of beam-loading in the final energy spread. Self-focusing is the key parameter to trigger upstream injection. 

\section{Conclusion and opening}

Starting from a robust plasma target design composed of two chambers, with dopant mitigated in the first part, the method presented here allowed for the generation of a large number of electron beams satisfying the initial filter '$Q >$ 30~pC \& $E_{med} >$ 150~MeV \& $\delta{}E_{mad} <$ 5~\% \& $\epsilon_y <$ 2~$\mu$m' with origins from different input configurations (LPI working points). This was allowed by fast simulations with the SMILEI code combined to computing time allocated by GENCI at TGCC. 

Beams matching the filter corresponded to a laser focus in vacuum placed at the end of the accelerating chamber. Plasma self-focusing allowed for an earlier injection and longer accelerating distance (typically all along chamber 2).

The divergence was outside of the scope but previous works proved the efficiency of plasma outramp to deal with this issue. This work could be done as a post-process of the present results.


All present results are left open to the scientific community, so that any researcher may use them to find optimal working points for a specific LPI, even including artificial intelligence and neural network studies.

\begin{acknowledgments}
This work was granted access to the HPC resources of TGCC under the allocations 2021 - A0110510062 and 2022 - A0130510062 made by GENCI for the project Virtual Laplace.
\end{acknowledgments}



\nocite{*}
\bibliography{biblio.bib}

\end{document}